\documentclass[reprint,twocolumn,showpacs,amsmath,amssymb]{revtex4-1}
\usepackage{bm}
\usepackage{hyperref}
\usepackage{tikz}

\begin{document}

\date{\today}

\title{Collective modes in a Dirac insulator with short range interactions}

\author{Xi Luo$^{1}$, Yue Yu$^{1,2}$}

\author{ Long Liang$^{2}$}
\email{lliang@itp.ac.cn}

\affiliation {${}^1$Department of Physics, Center for Field
Theory and Particle Physics, State Key Laboratory of Surface Physics and Collaborative Innovation Center of
 Advanced Microstructures, Fudan University, Shanghai 200433,
China \\
${}^2$State Key Laboratory of Theoretical Physics, Institute of
Theoretical Physics, Chinese Academy of Sciences, P.O. Box 2735,
Beijing 100190, China }

\date{\today}

\begin{abstract}

We study a Haldane model with nearest neighbor interactions. We find one-dimensional like collective modes arising due to the interplay of pseudo-spin and valley degrees of freedom. In the large band gap or moderate interaction limit, these excitations are low energy modes lying in the band gap. The dispersion relations are qualitatively different  in trivial insulator  phase and Chern insulator phase, thus can be used to identify the  topology of the Haldane model  with the bulk property. We also discuss how to detect these modes in  cold atom systems.  An abelian gauge theory will emerge when a physical current-current interaction is introduced to the Haldane model or the Kane-Mele model.

 \end{abstract}

\maketitle

\section{Introduction }

The discovery of graphene \cite{gn} opens the door of studying Dirac fermions in condensed matter systems. Topological insulator \cite{ti1,ti2} is a novel state of matter hosting gappless Dirac fermions on the surface and can be described by a massive Dirac equation in the bulk.
The collective excitations in these systems attract much interests both theoretically and experimentally.  The plasmon in graphene under various conditions has been discussed extensively \cite{va,wun,hw,gan,gon,pya,sarma,ande,sode,stefan} and observed experimentally \cite{nature1}.  Plasmons are collective modes that are usually absent in an insulator.
It's shown that  due to  the interplay between Dirac and Schr\"{o}dinger fermions \cite{stefan}, plasmons appear in 2D topological insulator, e.g., Hg(Cd)Te quantum wells.  These modes are also sensitive to the bulk topology \cite{stefan,zhou}. The spin-momentum locking effect on the surface of a 3D strong topological insulator induces a curious effect: density fluctuations induce spin fluctuations and vice versa and this gives the spin-plasmon \cite{zhang}. The observation of Dirac plasmons  in topological insulators  is reported recently \cite{natp1}. Ref. \cite{stauber} reviews the recent progress of plasmons in graphene and topological insulator.

In this paper we study the collective modes in a Haldane model \cite{hald}, which has been realized in the optical lattice experiments very recently \cite{1406}. It is natural to consider short range interactions in the optical lattice, so the collective modes are analogous with the zero sound in a neutral fermi liquid. Plasmons in gapped graphene have been studied in Ref. \cite{pya} where the density-density correlation is calculated.    Surprisingly, we find  the interplay of pseudo-spin and valley degrees of freedom gives a new phenomenon: one dimensional-like collective modes that are not reported before emerge.
 The underline lattice $C_3$ symmetry is broken by the quantum anomaly. These modes are qualitatively different in topological trivial phase and non-trivial phase and can be used to identify the topology of Haldane model.

Random phase approximation (RPA) is widely used to study the collective modes and its validity is established in graphene \cite{sarmar}. Here we use a different form of RPA \cite{nagaosa}, i.e., we derive the one-loop effective field theory of the collective modes. Around a single valley, the effective theory is equivalent to a  Maxwell-Chern-Simons theory \cite{pp,frad,kondo}.
In \cite{pp} the authors eliminated a valley with larger gap. However, we  find the contribution of the heavier Dirac point is as large as the lighter one up to the lowest order \cite{works}. If we consider the contributions of  both valleys properly, the effective theory is no longer equivalent to a Maxwell-Chern-Simons theory.  Combining the contributions from two Dirac points, there is no gauge symmetry anymore, while new effects arise.
 We find one-dimensional collective modes  emerge in both topological trivial (trivial insulator) and topological nontrivial (Chern insulator) cases. The behavior of the collective modes is qualitatively different: The  wave vector is perpendicular to the direction of the two Dirac points in a trivial insulator while it is along the direction of the two Dirac points in a Chern insulator. The gaps of the modes are also different.  When the band gap is large and the interaction is not too small, both modes can lie in the band gap. The  in-gap collective modes have been observed in the topological Kondo insulator $\mathrm{SmB_6}$ \cite{smb_e}. These modes  originate from the strong interactions \cite{smb_t}. In our case, the in-gap modes appear for weak or moderate interactions and strong enough  interactions will induce a Mott transition \cite{mottran}.

Recently, the quantum simulation of the dynamic gauge theories attracts a lot of interests in the cold atom and other systems \cite{buch,kapit,zohar,zoharc,ban,zoharcr,mac,tag,bane,zoharprl,tagl,zohara,wiese}.   Most researches focused on simulating  the lattice gauge theories \cite{hon,orl,chand}.  Ref. \cite{pp}
provides  a new idea although, in our opinion, the authors missed  the effect of the heavier Dirac  fermion. We thus construct a model in which the gauge theories do emerge. Taking account of the rapid development of cold atom experiments, we hope the model can be realized in the near future.

The rest of this paper is organized as follows. In Sec. \ref{section1} we introduce the model and derive the low energy effective  Lagrangian of the model. In Sec. \ref{section2} we  present the dispersion relations of the collective modes in both trivial insulator and Chern insulator phases. We then discuss how to detect them in cold atom systems. A different interaction is needed to get an emergent gauge theory and we briefly discuss this in Sec. \ref{section3}.  Finally, a conclusion is given in Sec. \ref{section4}.

\section{Model and method}\label{section1}

\begin{figure}
\includegraphics[width=0.3\textwidth]{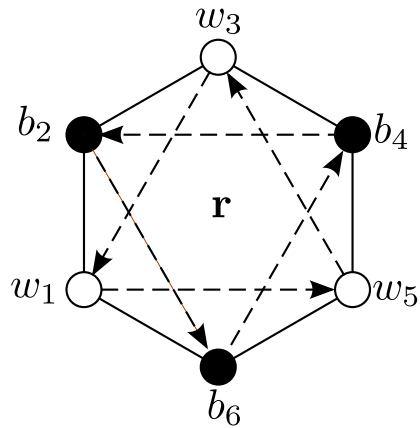}
\caption{\label{lattice} Haldane model on honeycomb lattice. Black and white points  mark the sublattice. Arrows  denote the positive phase hopping. {\bf r} is the center of the plaquette.}
\end{figure}

The Haldane model on honeycomb lattice \cite{hald, 1406} is governed by the following Hamiltonian:
\begin{equation}
H_0=-\sum_{\langle ij \rangle_1}t_{ij} c^\dagger_{i}c_{j}-\sum_{\langle ij \rangle_2}\lambda_{ij}e^{i\phi_{ij}}c_{i}^\dagger c_{j}+\Delta\sum_{i}(-1)^ic^{\dag}_{i}c_{i},\label{hald}
\end{equation}
where $c^{\dag}_{i}$ ($c_i$) is the fermion creation (annihilation) operator on site $i$,  $t_{ij}$  is the nearest hopping amplitude and $\lambda_{ij}$ is the next-to-nearest neighbor one. $\phi_{ij}$ is  the directed phase defined along the arrows in Fig.\ref{lattice} and  when $|\phi_{ij}|\ne 0,\pi$,  it will break the time reversal symmetry.  $2\Delta$ is the onsite energy difference of sublattices and it breaks inversion symmetry. Time reversal symmetry breaking and inversion symmetry breaking terms both give a band gap. However, the gaps are different in topological nature.  As a result, varying $\Delta/\lambda$  induces a topological phase transition from a Chern insulator to a trivial insulator \cite{hald}.  Although important, Haldane model seems unrealistic in materials. Fortunately, the rapid progress in cold atom experimental techniques gives new possibilities.  In a recent  experiment \cite{1406}, Haldane model has been realized in  a time-modulated optical lattice. Experimentally, $t_{ij}$ and $\lambda_{ij}$ are bond-dependent and $\phi_{ij}$ is controlled by the modulation phase. Our results don't rely on the details of these parameters and  we will take $t_{ij}=t$, $\lambda_{ij}=\lambda$ and $|\phi_{ij}|=\pi/2$ for simplification.

There are 3 nearest neighbors on the honeycomb lattice (set the lattice constant to be $1$),
\begin{equation}
\mathbf{b}_1=(\frac{1}{2\sqrt{3}},\frac{1}{2}), \mathbf{b}_2=(\frac{1}{2\sqrt{3}},-\frac{1}{2}), \mathbf{b}_3=(-\frac{1}{\sqrt{3}},0),
\end{equation}
and 6 next-to-nearest neighbors,
\begin{eqnarray}
\mathbf {a}_1=(\frac{\sqrt{3}}{2},-\frac{1}{2}), \mathbf {a}_2=(0,1), \mathbf {a}_3=(\frac{\sqrt{3}}{2},\frac{1}{2}),&&\nonumber\\
\mathbf{a}_4=(-\frac{\sqrt{3}}{2},\frac{1}{2}), \mathbf {a}_5=(0,-1), \mathbf {a}_6=(-\frac{\sqrt{3}}{2},-\frac{1}{2}).&&
\end{eqnarray}
Now we rewrite the Hamiltonian $H_{0}$ in the momentum space,
\begin{eqnarray}\label{haldk}
&&H_0=-t\sum_{\mathbf{k}}((\sum_{i=1}^3 e^{i\bf{k}\cdot\mathbf{b}_i})b_{\mathbf{k}}^\dagger w_{\mathbf{k}}+h.c.)\nonumber\\
&&+\sum_{\bf{k}}[\Delta-2\lambda
(\sin(\sqrt{3}k_x/2+k_y/2)-\sin(k_y)-\nonumber\\
&&\sin(\sqrt{3}k_x/2-k_y/2)
)](b^\dagger_{\mathbf{k}}b_{\mathbf{k}}-w^\dagger_{\mathbf{k}}w_{\mathbf{k}}).
\end{eqnarray}
where we label the two sublattices by $b$ and $w$.
Now let us consider the low energy behavior around the two independent Dirac points $\vec{K}$ and $-\vec{K}$, with
$\vec{K}=(0,\frac{4\pi}{3})$.
Here we take $\hbar=1$. The mass gap at $K_\pm$ is $m_{\pm}c^2=\Delta\pm 3\sqrt{3}\lambda$
 and $c=\frac{\sqrt{3}}{2}t$ is the Fermi velocity and we take $c=1$ in the following;
 At half-filling, the Fermi level  lies in the gap.  In the long wavelength limit, Eq. (\ref{haldk}) can be approximated around $K_\pm$ to the linear order of the momentum
 $p$
 \begin{eqnarray}
 H_0\approx\sum_{p,v} \psi^\dag_{p,v}{\cal H}_0(p,v)\psi_{p,v},
 \end{eqnarray}
  where $v=\pm$ is the valley index labeling the Dirac points $K_\pm$, $\psi_{p, v}=(b_{p v},w_{p v})^T$ is the two component fermion spinor near the Dirac point $K_v$ and $b,w$ label the two sub-lattices. The Hamiltonian ${\cal H}_0(p, v)$ is a $2\times 2$ matrix given by
\begin{eqnarray}\label{h0}
{\cal H}_0(p,v)= v  p_y\tau^x-   p_x\tau^y-m_{v} \tau^z.
\end{eqnarray}
Notice the $v$ before $p_y$. Because of this sign, the chiralities of the Dirac points are the same if  the signs of $m_\pm$ are different. This gives a simple way to calculate
the Chern number: $\nu=\frac{1}{2}[{\rm sgn}(m_{+})-{\rm sgn}(m_{-})]$. When $\nu=0$, the system is a trivial insulator; when $\nu=\pm 1$, it's a Chern insulator. As we will show below, $v$ also plays an important role in the effective theory of collective modes.

We consider the nearest neighbor Hubbard interaction
\begin{equation}
H_{int}=U\sum_{\langle ij\rangle_1}n_in_j.\label{int}
\end{equation}
In order to take the correct continuum limit, we rewrite the interaction Eq. (\ref{int}) by   summing over the plaquettes (see Fig. \ref{lattice}), i.e.,
\begin{eqnarray}
H_{int}&=&\frac{U}2
\sum_{\langle bw\rangle\in \rm plaq}(n_{w_1}n_{b_2}+n_{b_2}n_{w_3}+n_{w_3}n_{b_4}\nonumber\\
&&+n_{b_4}n_{w_5}+n_{w_5}n_{b_6}+n_{b_6}n_{w_1}).\label{40}
\end{eqnarray}
Now we approximate $n_{i}$ by its value at the center of the plaquette (denote as ${\bf r}$), for example,
\begin{eqnarray}
n_{w_1}&\approx&n_w({\bf r})-\frac{\sqrt{3}}{3}(\frac{\sqrt{3}}{2}\partial_x n_w({\bf r})+\frac{1}{2}\partial_y n_w({\bf r})),\\
n_{b_2}&\approx&n_b({\bf r})-\frac{\sqrt{3}}{3}(\frac{\sqrt{3}}{2}\partial_x n_b({\bf r})-\frac{1}{2}\partial_y n_b({\bf r})).
\end{eqnarray}
Therefore after some arithmetics, Eq. (\ref{int}) becomes,
\begin{equation}\label{lint}
H_{int}=\frac{g^2}2
\sum_{{\bf r}}(6n_{b}({\bf r})n_{w}({\bf r})+\frac{\sqrt{3}}{2}\nabla n_{b}({\bf r})\cdot\nabla n_{w}({\bf r})),
\end{equation}
where $g^2=U$. In the continuum limit, we drop the second term. The disappearance of the first order terms is guaranteed by the lattice symmetry.

To make further progress we write the density-density interaction in the suggestive way:
\begin{eqnarray}
6n_{b}n_{w}=(n_{b}+n_{w})^2-(c^{\dag}_{b}c_{w}+c^{\dag}_{w}c_{b})^2-(-ic^{\dag}_{b}c_{w}+ic^{\dag}_{w}c_{b})^2.\nonumber
\end{eqnarray}
We have omitted the $n_b+n_w$ terms since they can be absorbed into chemical potential.

Now we decouple the interaction Eq. (\ref{lint}) using the following Hubbard-Stratonovich transformation:
\begin{eqnarray}
&&\exp\biggl\{-\frac{ig^2}{2}\int dt d^2r~ 6n_{b}(\mathbf{r})n_{w}(\mathbf{r})\biggr\} \nonumber\\
&&=\int\mathcal {D} a_{\mu} \exp\biggl\{i\int dtd^2r (\frac{1}{2}a_{\mu} a^\mu +gj_{\mu} a^\mu)\biggr\},\label{intchal}
\end{eqnarray}
with  $\mu=0,1,2$ and
$j_0=n_{b}+n_{w}$, $j_1=-ic^{\dag}_{b}c_{w}+ic^{\dag}_{w}c_{b}$ and $j_2=-c^{\dag}_{b}c_{w}-c^{\dag}_{w}c_{b}$. In Eq. (\ref{intchal}) we use Einstein's summation  convention and the metric is $g^{\mu\nu}=diag(1,-1,-1)$.
We then approximate the 3-`current' by $j_{\mu}\approx \sum_v j_{\mu v}$ with $j_{\mu v}$ defined near $K_v$, i.e.,
\begin{eqnarray}\label{curr}
&&j_{0 v}=\psi^\dag_{v}({\bf r})\psi_{v}({\bf r}), \nonumber\\
&&j_{1 v}=\psi^\dag_{v}({\bf r})\tau^y\psi_{v}({\bf r}), \\
&&j_{2 v}=-\psi^\dag_{v}({\bf r})\tau^x\psi_{v}({\bf r}). \nonumber
\end{eqnarray}
It's crucial to notice that the  `current' $j_2=j_{2,+}+j_{2,-}$ is not the physical current. The physical current should be $j_{2,phy}=j_{2,+}-j_{2,-}$ because the coefficient of $p_y$ has  an opposite sign around the two Dirac points, see Eq. (\ref{h0}). This simple fact leads to  some interesting physical phenomena.

Now we can write  down the low energy effective Lagrangian  as:
\begin{eqnarray}
{\cal L}&=&\sum_v(\bar\psi_{v}(t,{\bf r})(i {\not}\partial_{v} +m_{v})\psi_{v}(t,{\bf r})\nonumber\\
&&+g j_{\mu} a^\mu+\frac{1}{2}a_{\mu} a^\mu ),\label{lwll}
\end{eqnarray}
where $\bar\psi_{v}=\psi^\dag_{v}\gamma^0_v$ and  ${\not}\partial_{v}=\gamma^{\mu}_v\partial_{\mu}$,
 $\gamma^0_v=\tau^z$, $\gamma^1_v=i\tau^x$, $\gamma^2_v=vi\tau^y$, $\partial_{\mu}=(\partial_{t},\partial_{x},\partial_{y})$.  We can rewrite the `current' in terms of the gamma matrices:  $j_v^{0/1}=\bar{\psi}_{v}\gamma_v^{0/1}\psi_{v}$ and $j^{2}_v=v\bar{\psi}_{v}\gamma_v^{2}\psi_{v}$.  Notice that the effective Lagrangian (\ref{lwll}) is rotationally invariant. This is the result of the long wavelength limit. If the higher orders beyond the linear approximation of the wave vector are included, the symmetry should be reduced to the lattice $C_3$ point group symmetry.

\section{collective modes and experimental implications}\label{section2}

The physical  meanings of $a_0$, $a_x$ and $a_y$ are the charge and pseudo-spin densities.  We don't decouple in $s^z$ channel because the density fluctuations only  couple to $s^x$ and $s^y$  components of pseudo-spin \cite{zhang}. To discuss the properties of the collective modes we integrate over the fermions. The one-loop result has the form  $\mathcal{L}_{\mathrm{eff}}=\frac{1}{2}(a^2-g^2 a^{\mu}\Pi_{\mu\nu}a^{\nu})$.  $\Pi_{\mu\nu}$ is nothing but the bare susceptibility tensor, this verifies the equivalence between our approach and  RPA.

Up to $1/|m_{\pm}|$ order, the one-loop result is given by (see Appendix. \ref{app1}) 

\begin{eqnarray}\label{leff}
\mathcal{L}_{\mathrm{eff}}&=&\frac{g^2}{8\pi}\mathrm{sgn}(m_+)\varepsilon^{\mu\nu\rho}a_{\mu}\partial_{\nu}a_{\rho}-\frac{g^2}{24\pi|m_+|}F^{\mu\nu}_aF_{a ,\mu\nu}\nonumber\\
&&-\frac{g^2}{8\pi}\mathrm{sgn}(m_-)\varepsilon^{\mu\nu\rho}b_{\mu}\partial_{\nu}b_{\rho}-\frac{g^2}{24\pi|m_-|}F^{\mu\nu}_bF_{b ,\mu\nu}\nonumber\\
&&+\frac{1}{2}a^2,
\end{eqnarray}
where $F_{a,\mu\nu}=\partial_{\mu}a_{\nu}-\partial_{\nu}a_{\mu}$, $F_{b,\mu\nu}=\partial_{\mu}b_{\nu}-\partial_{\nu}b_{\mu}$ and $b_0=-a_0$, $b_1=-a_1$ and $b_2=a_2$.
The first line of Eq. (\ref{leff}) is the contribution of $K_+$ and the second line comes from $K_-$.  Two Chern-Sinoms terms corresponding to two Dirac points come from the $m_\pm$-zeroth order contributions and are the quantum anomalies of the Dirac fermion. We emphasize $a_\mu$ and $b_\mu$ are not independent and will discuss its physical results soon.

We make some observations about  the contribution of a single Dirac point:  There is  a constraint of $a_{\mu}$ fields which is a direct result of the equation of motion. The physical reason of the constraint is  that the charge density only couples to the transverse component of spin density \cite{zhang}. The effective action around a single Dirac point becomes a Maxwell-Chern-Simons theory after a dual transformation \cite{pp,frad,kondo}. In \cite{pp}, the authors eliminate the Dirac point with  a larger mass.
 However, up to the lowest order,  the contribution of  a Dirac point only dependents on the sign of  its mass (see Eq. (\ref{leff})), thus there is no way to get rid of the effect of the heavier Dirac fermion, as emphasized by Haldane \cite{hald}.  Notice that $a_\mu$ and $b_\mu$ have the same magnitude but the sign may be the same or opposite.  This leads to two remarkable results: (i)
 The rotational symmetry is broken by the quantum anomaly in the one loop approximation.  This means that the underline lattice $C_3$ symmetry is also broken if the high order expansions are considered.  (ii)
 The effective theory is no longer a gauge theory because of the cancellation of some of the components of the $a_\mu$ field. However, this also means that the $a_\mu$ field cannot be completely canceled even in the topological trivial case.   In the following we study the dispersions of the collective modes $a_\mu$ in the topological trivial and nontrivial phases.

For $\mathrm{sgn}(m_+)=\mathrm{sgn}(m_-)$, i.e., the trivial insulator case, the collective modes are dynamical in the lowest order approximation so we can omit the $1/|m_{\pm}|$ corrections. The pole of Green's function of $a_{\mu}$ gives the following dispersion relation:
\begin{eqnarray}\label{dis1}
\omega^{2}=p_x^2+m_A^2,
\end{eqnarray}
where $m_A=2\pi/g^2$.  This dispersion is one-dimensional like. $p_x$ is the direction perpendicular to the vector  that connects the two Dirac points, so it's physically distinct from $p_y$. When $m_A<\mathrm{min}(|m_+|,|m_-|)$, this mode lies in the band gap and will not decay to single particle excitations. Let the band gap be $\Delta_{t}$ and recover the `speed of  light' $c$, the in-gap condition becomes $U\Delta_t>2\pi c^2$ which can be fulfilled in the large band gap or  the strong interaction limit.  A strong enough interaction will induce a Mott transition \cite{mottran}. However, on the one hand, the critical interaction of Mott transition is of the order of band width \cite{mottran}. On the other hand,  the band   gap can be tuned in cold atom systems. We thus expect that the in-gap modes can be observed experimentally.

For $\mathrm{sgn}(m_+)=-\mathrm{sgn}(m_-)$, i.e., the Chern insulator case, to get the dispersion of the collective modes one have to consider the $1/|m_{\pm}|$ corrections. We then find two dispersions around the Dirac points
\begin{eqnarray}
&&\omega^2\approx 6\pi M/g^2+ p_x^2+\beta p_y^2,\\
&&\omega^2\approx 6\pi M/g^2+\alpha g^2 M p_y^2+p^2_x,\label{sec}
\end{eqnarray}
with $M =\frac{|m_+ m_-|}{|m_+|+|m_-|}$, and $\alpha, \beta$ are dependent on $|m_+|, |m_-|$  and are of order 1. These modes can also lie in the band gap. Let the band gap be $\Delta_n$ and $|m_+|=|m_-|$, the in-gap condition is $U\Delta_{n}>3\pi c^2$. Comparing this condition with the one in topological trivial case, we find that for the same $U$ and $c$, the band gap $\Delta_n$ should be larger than $\Delta_t$ to get the in-gap modes. Generally speaking, this means that the gap of the in-gap collective modes in a Chern insulator is larger than that in a trivial insulator.  Since we are considering the large band gap limit, $M>>1$, the dispersion Eq. (\ref{sec}) is one-dimensional like and can be approximated as
\begin{eqnarray}\label{dis2}
\omega^2=6\pi M/g^2+\alpha g^2 M p_y^2.
\end{eqnarray}
This mode is the characteristic mode in the Chern insulator phase: it's one-dimensional like with large velocity.

Now we get the main results of this paper:  we find that in an insulator with Dirac points, the lattice point
group symmetry may be broken by the quantum anomaly and there are one-dimensional like collective modes. These bulk modes also reveal the topology of the insulator: in the large band gap limit, the gap of collective modes of a Chern insulator is larger than that of a trivial insulator and both modes can lie in the band gap.
What is more, the dispersion of the collective modes in a Chern insulator is along  a given direction, say, connecting the two Dirac points while in a trivial insulator it is perpendicular to that direction.  We are not aware of collective modes with these properties.

Our predictions can be examined in cold atom experiments.  The collective modes can by measured  through Bragg scattering. In the Bragg scattering process,  two laser beams with 3-momentum difference  $q_\mu=k_{1\mu}-k_{2\mu}$ are sent to  the trapped cold atom gases on the optical lattice. The illuminated cold atom system responds to this perturbation with a density fluctuation
 $H'= (V/2)(\delta j_{0}({\bf q})e^{iq_0t}+h.c.)$.
   The atoms then absorb a momentum after  the Bragg scattering. The rate of the momentum transfer can be measured by the time of flight images and  be related to the dynamic structure factor  by $d {\bf P}/dt
   =2\pi{\bf q}(V/2)^2[S(q_\mu)-S(-q_\mu)]\propto (1-e^{-\beta q_0})S(q_\mu)\propto \mathrm{Im} (\chi_{q})$ where ${\bf P}$ is the total momentum  of the atoms and $\chi_q$ is the density-density correlation \cite{bragg} .  Because the interaction is relatively weak, the heating arising from the interaction is negligible and then  $\partial {\bf P}/dt$ can be evaluated at zero temperature, i.e.,  $\propto S(q_\mu)$. It may be hard to detect the pseudospin density correlation directly, however,  the density correlation is enough for our purpose. In our case, the density-density correlation is just the correlation of $a_0$ and can be calculated easily.

   For a trivial insulator, $$\langle a_{0,-q}a_{0,q}\rangle=2\frac{\omega^2-m_A^2}{\omega^2-p_x^2-m_A^2},$$  with $m_A=2\pi/g^2$, this gives
\begin{eqnarray}\label{structure1}
 \mathrm{Im} (\chi_{q})=2\pi p^2_x\delta(\omega^2-p_x^2-m_A^2).
 \end{eqnarray}

 For a Chern insulator, we consider two  limits, i.e., $|m_{+}|=|m_{-}|\equiv M$ and $|m_{+}|>>|m_{-}|\equiv M$.

For $|m_+|=|m_-|\equiv M$, $a_y$ mode is decoupled from $a_0$ and $a_x$ and the dispersion of $a_y$ mode is $$\omega^2=m_{A,c}^2+p^2_x,$$ with $m_{A,c}^2=3\pi M/g^2$. This dispersion is along $p_x$ direction with a small  `velocity'. The one-dimensional nature of this  mode is due to the special parameter. What's more, this mode  cannot be detected by density-density correlation.
On the other hand, the characteristic mode can be revealed from the density-density correlation: $$\langle a_{0,-q}a_{0,q}\rangle=\frac{2m^2_{A,c}(\omega^2-p^2_y-m^2_{A,c})}{(\omega^2-E^2_{\mathbf{p}})(p^2_y+m^2_{A,c})},$$
where $E^2_{\mathbf{p}}=m^2_{A,c}+p^2_y+p^2_x+\frac{1}{4\pi^2}\frac{m^4_{A,c}g^4}{m^2_{A,c}+p^2_y}p^2_y\approx m^2_{A,c}+\frac{m^2_{A,c}g^4}{4\pi^2}p^2_y$, and we have
\begin{eqnarray}
\mathrm{Im}(\chi_{q})\approx \frac{g^4m^2_{A,c}}{2\pi}p^2_{y} \delta(\omega^2-E^2_{\mathbf{p}}).\label{structure2}
\end{eqnarray}

 For $|m_{+}|>>|m_-|\equiv M$, $$\langle a_{0,-q}a_{0,q}\rangle\approx\frac{2(\omega^2-m^2_{A,c})(\omega^2-p^2_x-p^2_y-m^2_{A,c})}{(\omega^2-p^2_x-m^2_{A,c})(\omega^2-m^2_{A,c}-\frac{m^2_{A,c}g^4}{4\pi^2}p^2_y)},$$
here $m^2_{A,c}=6\pi M/g^2$. There are two modes and are all one-dimensional like due to the special parameter. The characteristic mode is along the $p_y$ direction. Along  the $p_y$ direction the structure factor has the same form as Eq. (\ref{structure2}):
\begin{eqnarray}\label{structure3}
\mathrm{Im}(\chi_{p_x=0})\approx \frac{g^4 m^2_{A,c}}{2\pi}p^2_{y} \delta(\omega^2-m^2_{A,c}-\frac{m^2_{A,c}g^4}{4\pi^2}p^2_y).
\end{eqnarray}

 Finally, we would like to point out that although the $C_3$ symmetry is broken, the coordinate axes $x$ and $y$ are arbitrarily chosen. In practice, it should be pinned by various imperfects, impurities and other circumstances.


\section{emergent gauge theory}\label{section3}

In this section, we construct an interaction  such that gauge theory emerges.
As we have emphasized several times in the previous sections, the `current' $j_{2}=j_{2,+}+j_{2,-}$ is not the physical current. The physical current should be $$j_{2,phy}=j_{2,+}-j_{2,-}.$$ To get the emergent gauge field, we  simply replace $j_2$ by $j_{2,phy}$ in Eq. (\ref{lwll}), then the  model becomes a Thirring model \cite{thir, thirc}.  We construct a lattice interaction  in Appendix \ref{app2}.   After integrating over the fermions the effective Lagrangian becomes (See  Appendix \ref{app1}):
\begin{eqnarray}\label{gauge}
\mathcal{L}_{\mathrm{eff}}&=&\frac{g^2\nu}{4\pi} \varepsilon^{\mu\nu\rho}a_{\mu}\partial_{\nu}a_{\rho}-\frac{g^2}{24\pi M}F^{\mu\nu}F_{\mu\nu}
+\frac{1}{2}a^2,
\end{eqnarray}
 where $\nu=\frac{1}{2}(\mathrm{sgn}(m_+)-\mathrm{sgn}(m_-))$ is the Chern number, $M=\frac{|m_+ m_-|}{|m_+|+|m_-|}$ and $F_{\mu\nu}=\partial_{\mu}a_{\nu}-\partial_{\nu}a_{\mu}$. When $\nu=0$, this is a massive Maxwell theory (Proca theory); when $\nu\ne 0$, it's equivalent to a Chern-Simons-Maxwell theory after a dual transformation \cite{kondo}. The mass of Proca theory is $6\pi M/g^2$ while the mass of Chern-Simons-Maxwell theory is $2\pi/g^2+\mathcal{O}(1/M)$. The mass of the Proca theory can hardly be tuned small because  we require $M$ to be large to get the effective action. Hence,  it is impossible to discuss the confinement effect \cite{poly}. The mass of Chern-Simons-Maxwell theory can be tuned small if the interaction is large or the `speed of light' is small.

Kane-Mele model \cite{KM,KM1} is essentially two copies of Haldane model related by time reversal symmetry and  can also be realized by the same technique in Ref. \cite{1406}. The collective modes discussed in Sec. \ref{section2} also appear in Kane-Mele model with short  range interactions \cite{KMH}
 such as $U\sum_{\langle ij\rangle_1;\sigma}n_{i\sigma}n_{j
\sigma}$. What is more, it is possible to  reach to the confinement limit if the interaction is of {\it physical current-current type}. In this case, { due to introducing of the spin degrees of freedom}, we  get a mutual Chern-Simons theory (up to $\mathcal{O}(1/|m_{\pm}|)$):
\begin{eqnarray}\label{mutualc}
\mathcal{L}_{\mathrm{eff}}&=&\frac{g^2\nu_{\sigma}}{4\pi} \varepsilon^{\mu\nu\rho}a_{\sigma \mu}\partial_{\nu}a_{\sigma\rho}
+\frac{1}{2}a^2_{\sigma},
\end{eqnarray}
where $a_{\sigma}$ is the collective modes of spin-$\sigma$ fermions and $\nu_{\uparrow}=-\nu_{\downarrow}$, as is required by time reversal symmetry. Let $A_{\mu}=a_{\mu\uparrow}-a_{\mu\downarrow}$, In terms of $A_{\mu}$ and $a_{\mu\uparrow}$, the Lagrangian ${\cal L}_{\mathrm{eff}}$ becomes,
\begin{eqnarray}
\mathcal{L}_{\mathrm{eff}}&=&\frac{1}{2}a^\mu_\uparrow a_{\mu\uparrow}+\frac{1}{2}(\frac{8\pi ^2 }{\alpha^2}A^2-\frac{4\sqrt{2}\pi}{\alpha}A_\mu a^\mu_\uparrow+a^\mu_\uparrow a_{\mu\uparrow})\nonumber\\
&&-\sqrt{2}\epsilon^{\mu\nu\rho} a_{\mu\uparrow} \partial_\nu A_\rho +\frac{2\pi}{\alpha}\epsilon^{\mu\nu\rho} A_\mu \partial_\nu A_\rho.
\end{eqnarray}
Now let us integrate out the $a_{\mu\uparrow}$ field, we get the effective Lagrangian,
\begin{equation}
{\cal L}_{eff}= -\frac{F^{\mu\nu} F_{\mu\nu}}{4}+\frac{1}{2}m_A^2 A^\mu A_\mu,
\end{equation}
 with $F_{\mu\nu} = \partial_{\mu} A_{\nu}-\partial_{\nu} A_{\mu}$ and $m_A = 2\pi/g^2$. In this case, it is possible to pursue the massless Maxwell limit and study the confinement effect in the continuum gauge theory experimentally.

 Notice that the rotational symmetry is kept for the physical current-current  interactions.

\section{ Conclusions }\label{section4}

We have investigated the collective modes in a Dirac insulator, i.e., Haldane model with short  range interaction. We  have shown one-dimensional excitations emerge due to the interplay between the valley and  the pseudo-spin degrees of freedom.  For those linear excitations, the $C_3$ symmetry is broken by quantum anomaly. The qualitative behavior of the modes is quite different in topological trivial  phase and nontrivial phase thus can be used to identify the phase of the Haldane model  with its bulk property.
 We also  discuss how to detect these modes in cold atom experiments. The interplay between Dirac  fermions and Schr\"odinger fermions may lead to new phenomena and  will be studied in a future work, where the effects of lattice will be considered as well. We also construct  the interactions that lead to emergent gauge theories of the collective modes, e.g., Proca theory and Chern-Simons-Maxwell theory.

\acknowledgments
The authors thank  Yuanpei Lan, Ziqiang Wang and Yong-Shi Wu for helpful discussions.  We also thank an
anomalous referee for drawing our attentions to the collective modes of the model. This work is supported by the 973 program of MOST of China (2012CB821402), NNSF of China (11174298, 11121403, 11474061).

\appendix

\section{Effective field theory of collective modes}\label{app1}

In this appendix we show how to get Eq. (\ref{leff}) from Eq. (\ref{lwll}).

\begin{eqnarray}
\mathcal{L}=\mathcal{L}_{+}+\mathcal{L}_{-}+\frac{1}{2}a^2,\nonumber
\end{eqnarray}
where $\mathcal{L}_{\pm}$ is the  Lagrangian around Dirac point $K_{\pm}$:

$\mathcal{L_+}=\bar{\psi}_{+}(i\gamma^0\partial_0+i\gamma^1\partial_1+i\gamma^2\partial_2+m_{+})\psi_{+}
+g a_{\mu}\bar{\psi_+}\gamma^{\mu}\psi_{+},$

$\mathcal{L_-}=\bar{\psi}_{-}(i\gamma^0\partial_0+i\gamma^1\partial_1-i\gamma^2\partial_2+m_{-})\psi_{-}
+ga_{\mu}\bar{\psi_-}\gamma^{\mu}\psi_{-}.$

Note that the fermions around the two Dirac points are decoupled and can be integrated  out independently.
We can integrate over $\psi_+$ and the result is well known \cite{frad,kondo,jackiw}, in large mass limit we get the first line of Eq. (\ref{leff})

For the $\psi_-$ fermions, let $\gamma'^0=-\gamma^0$, $\gamma'^1=-\gamma^1$, $\gamma'^2=\gamma^2$, $b^0=-a^0$,  $b^1=-a^1$,  $b^2=a^2$,
 then $\mathcal{L}_-$ can be written as

$\mathcal{L_-}=-\bar{\psi}_{-}(i\gamma'^\mu\partial_\mu-m_{-})\psi_{-}+g b_{\mu}\bar{\psi}_-\gamma'^{\mu}\psi_{-}.$

Integrating over $\psi_-$,  $\mathcal{L}_-$ gives the second line of Eq. (\ref{leff}).

To get a gauge theory, $a_2$ should couple to the physical current $j_{2,+}-j_{2,-}$.  $\mathcal{L}_{+}$ is unchanged while $\mathcal{L}_{-}$ becomes:

\begin{eqnarray}
\mathcal{L_-}&=&\bar{\psi}_{-}(i\gamma^0\partial_0+i\gamma^1\partial_1-i\gamma^2\partial_2+m_{-})\psi_{-}\nonumber\\
&&+ga_{0}\bar{\psi}_-\gamma^{0}\psi_{-}+ga_{1}\bar{\psi}_-\gamma^{1}\psi_{-}-ga_{2}\bar{\psi_-}\gamma^{2}\psi_{-}\nonumber\\
&=&-\bar{\psi}_{-}(i\gamma'^\mu\partial_\mu-m_{-})\psi_{-}-g a_{\mu}\bar{\psi}_-\gamma'^{\mu}\psi_{-}.\nonumber
\end{eqnarray}

Integrating over $\psi_+$ and $\psi_-$ we get Eq. (\ref{gauge}).

\section{Current-current interaction on lattice}\label{app2}

To get the effective gauge theory, the interaction terms should be:

\begin{eqnarray}
&&a_{0}(\bar{\psi}_{+}\gamma^0\psi_{+}+\bar{\psi}_{-}\gamma^0\psi_{-})+a_{1}(\bar{\psi}_{+}\gamma^1\psi_{+}+\bar{\psi}_{-}\gamma^1\psi_{-})+\nonumber\\
&&a_{2}(\bar{\psi}_{+}\gamma^2\psi_{+}-\bar{\psi}_{-}\gamma^2\psi_{-}).\nonumber
\end{eqnarray}
As we have shown in the main text, the  first term comes from the nearest neighbor interaction $n_{i}n_{j}$.
To get a lattice version of the  last two terms, we first regularize  the second term as, e.g.:
\begin{eqnarray}
&&a_{1}(\bar{\psi}_{+}\gamma^1\psi_{+}+\bar{\psi}_{-}\gamma^1\psi_{-})=
a_1(\psi^{\dag}_{+}\sigma^y\psi_{+}+\psi^{\dag}_{-}\sigma^y\psi_{-})\nonumber\\
&&\sim
 a_{1,\mathbf{q}}\cos(k_1)\psi^{\dag}_{\mathbf{k}}\sigma^y\psi_{\mathbf{k-q}}\nonumber\\
&&=a_{1,\mathbf{q}}\cos(k_1)(-ic^{\dag}_{b,\mathbf{k}}c_{w,\mathbf{k-q}}+ic^{\dag}_{w,\mathbf{k}}c_{b,\mathbf{k-q}})\nonumber\\
&&=\frac{a_{1,\mathbf{q}}}{2}\sum_{i} (-ic^{\dag}_{b,i}c_{w,i\pm \mathbf{a}_1}
+ic^{\dag}_{w,i}c_{b,i\pm \mathbf{a}_1})e^{i\mathbf{q}\cdot(\mathbf{R}_i\pm \mathbf{a}_1)},\nonumber
\end{eqnarray}
where $i$ denotes the unit cell and $\mathbf{a}_1$ is one of the six next-to-nearest neighbors. And similarly for the third term,
\begin{eqnarray}
&&a_{2}(\bar{\psi}_{+}\gamma^2\psi_{+}-\bar{\psi}_{-}\gamma^2\psi_{-})=
a_2(\psi^{\dag}_{+}\sigma^x\psi_{+}-\psi^{\dag}_{-}\sigma^x\psi_{-})\nonumber\\
&&\sim
 a_{2,\mathbf{q}}\sin(k_1)\psi^{\dag}_{\mathbf{k}}\sigma^x\psi_{\mathbf{k-q}}\nonumber\\
&&=a_{2,\mathbf{q}}\sin(k_1)(c^{\dag}_{b,\mathbf{k}}c_{w,\mathbf{k-q}}+c^{\dag}_{w,\mathbf{k}}c_{b,\mathbf{k-q}})\nonumber\\
&&=\frac{a_{2,\mathbf{q}}}{2i}\sum_{i}\pm (c^{\dag}_{b,i}c_{w,i\pm \mathbf{a}_1}
+c^{\dag}_{w,i}c_{b,i\pm \mathbf{a}_1})e^{i\mathbf{q}\cdot(\mathbf{R}_i\pm \mathbf{a}_1)}.\nonumber
\end{eqnarray}

 Integrating over $a_{1}$ and $a_{2}$ fields, we get  the corresponding lattice interaction:

\begin{eqnarray}
&&c^{\dag}_{b,i}c_{b,i+\mathbf{a}_1}c^{\dag}_{w,i+\mathbf{a}_1}c_{w,i}+h.c..\nonumber
\end{eqnarray}
There are different ways to regularize the current $j_{2,phy}$ and the resultant lattice interactions are also different.

\end{document}